\documentclass[aps,prl,superscriptaddress,onecolumn,floatfix]{revtex4-1}

\usepackage{graphicx,color}
\usepackage{dcolumn}
\usepackage{bm}
\usepackage{stmaryrd}
\usepackage{latexsym}
\usepackage{amssymb}
\usepackage{amsfonts}
\usepackage{amsmath}
\usepackage{subfigure}
\usepackage{hyperref}
\usepackage{verbatim}
\usepackage{multirow}

\begin{document}

\preprint{APS/123-QED}

\title{Electronic nematicity in FeSe: a first-principles perspective}
\author{Xuanyu Long}
\affiliation{Institute for Advanced Study, Tsinghua University, Beijing 100084, China}
\author{Shunhong Zhang}
\affiliation{International Center for Quantum Design of Functional Materials (ICQD), Hefei National Laboratory for Physical Sciences at the Microscale, and Synergetic Innovation Center of Quantum Information and Quantum Physics, University of Science and Technology of China, Hefei, Anhui 230026, China}
\author{Fa Wang}
\affiliation{International Center for Quantum Materials, School of Physics, Peking University, Beijing 100871, China}
\affiliation{Collaborative Innovation Center of Quantum Matter, Beijing 100084, China}
\author{Zheng Liu}
\email{zheng-liu@tsinghua.edu.cn}
\affiliation{Institute for Advanced Study, Tsinghua University, Beijing 100084, China}
\affiliation{Collaborative Innovation Center of Quantum Matter, Beijing 100084, China}
\date{\today}

\begin{abstract}
Electronic nematicity is an important order in most iron-based superconductors, and FeSe represents a unique example, in which nematicity disentangles from spin ordering. It is commonly perceived that this property arises from strong electronic correlation, which can not be properly captured by density functional theory (DFT). Here, we show that by properly considering the paramagnetic condition and carefully searching the energy landscape with symmetry-preconditioned wavefunctions, two nematic solutions stand out at either the  DFT+$U$ or hybrid functional level,  both of which are lower in energy than the symmetric solution. The ground-state band structure and Fermi surface can be well compared with the recent experimental results.  Symmetry analysis assigns these two new solutions to the $B_{1g}$ and $E_u$ irreducible representations of the D$_{4h}$ point group.  While the $B_{1g}$ Ising nematicity has been  widely discussed in the context of vestigial stripe antiferromagnetic order, the two-component $E_u$ vector nematicity is beyond previous theoretical discussion. Distinct from the $B_{1g}$ order, the $E_u$ order features mixing of the Fe $d$-orbitals and inversion symmetry breaking, which lead to striking experimental consequences, e.g. missing of an electron pocket. 

\end{abstract}

\maketitle
\section{Introduction}
First-principles calculation within the framework of density function theory (DFT) has played an important role in understanding the iron-based superconductors. For the high-temperature paramagnetic phase, the standard local density approximation (LDA) or its generalized gradient approximation (GGA) extension can already qualitatively describe the Fermi surface topology and its orbital components~\cite{book12Dai,book15FeSc,rpp11ARPESrev}. For the low-temperature magnetic phase, the local spin density approximation (LSDA) or spin-polarized GGA (sGGA) can obtain the correct ground-state spin order in most cases~\cite{book12Dai,book15FeSc,nphys12magrev}. Additional corrections to the electronic correlation effects, e.g. the dynamical mean-field theory (DMFT), further reduce the quantitative discrepancies, such as the overestimation of the bandwidth and the magnetic moment~\cite{nmat11DMFT,nphys11DMFT}.

However, a first-principles description of the nematic order remains elusive. Dictated by a $C_4$ rotational symmetry breaking, the nematic phase remains as a paramagnetic metal~\cite{nphys14nem}. The attempt to explicitly break the $C_4$ symmetry by straining the lattice in the DFT simulation results in only negligible changes on the electronic structure~\cite{srep16strainDFT}, which is not surprising, since there is strong experimental evidence suggesting that the nematic order is of an electronic origin~\cite{sci12nemsus}. 

There are two main scenarios to explain the electronic mechanism of nematicity~\cite{nphys14nem}. One is to view it as a precursor of the stripe antiferromagnetic phase, in which the spin quadrupolar fluctuation first diverges before the ordering of the spins~\cite{prb08Kivelson,prb08Sachdev,prb12Fernandes}. The other is to invoke a spontaneous orbital ordering~\cite{prl09Ku,prb10OO,prb11Kontani}.  Despite very different microscopic origins, the formulated order parameters (OPs) all belong to the one-dimensional $B_{1g}$ irreducible representation (irrep) of the D$_{4h}$ point group, i.e Ising nematicity. By symmetry, these OPs are intertwined~\cite{prx17SOC}, and in reality electronic nematicity manifests in both the spin and the orbital sectors concurrently.

From the DFT perspective, there is no apparent way to take a composite spin order into account. Nevertheless, mature orbital-resolved approaches, e.g. DFT+$U$~\cite{jpc97LDA+U} and hybrid functional~\cite{jcp93hybrid}, allow probing potential instabilities in the orbital channel. FeSe represents an ideal platform to perform this numerical experiment, because no magnetic order has thus far been observed in bulk FeSe, unless high pressure is applied~\cite{prb12Mag,ncom16highp}. This provides a unique chance to study nematicity disentangled from spin ordering, and it is reasonable to speculate that in this case the orbital instability, if exists, can be detected by DFT+$U$ and hybrid functional calculations. In addition, inspired by the high superconducting transition temperature of monolayer FeSe on SrTiO$_3$ substrate~\cite{cpl12STO,nmat15STO}, extensive experimental data have been collected and crosschecked, in particular high-resolution angle-resolved photoemission spectroscopy (ARPES) data~\cite{prl14ARPESjp,prb14ARPESjp1,prb15ARPESjp2,prb15ARPESDing,prb15ARPESWatson,prb16ARPESWatson,njp17ARPESWatson,prb18ARPESKim,prl16ARPESFeng,prb16ARPESYan,prb16ARPESBrouet,srep16strainDFT,prb18ARPESBorisenko,prx18ARPESZhou,nmat16Liu,prx19Yi,comphy20Korea}, and thus systematic evaluation of the calculation results is possible.

The central finding of this article is that first-principles calculation can indeed provide important insights into the nematic electronic structure of FeSe. 
In previous theories, orbital splitting with opposite signs around $\Gamma$ and $M$ points is the dominant order parameter reproducing the ARPES observations~\cite{jpc15LiTao,prl16Kontani,prb18Chubukov}, but this order parameter cannot arise from mean-field treatment of local correlations on single Fe atoms, and sophisticated many-body theory has to be employed~\cite{prl16Kontani,prb18Chubukov}. Our results show that the important features of the nematic phase can be reproduced by careful treatment of local correlations within the state-of-the-art first-principles framework.

In the follows, we will first discuss the fundamental paramagnetic condition underlying the calculation, and then compare the obtained nematic band structure and Fermi surface with experimental results. The associated OPs are revealed by charge density, formulated by analytical representations  and classified by group theory, based on which we predict further experimental consequences. The technical details, in particular the importance of wavefunction preconditioning, are elaborated in Methods.


\section{Paramagnetic condition}\label{sec:special}

Within the nematic phase, FeSe is experimentally a paramagnetic metal without magnetic ordering. On the other hand, LSDA or sGGA has been shown to result in a quasi-degenerate lowest-energy manifold with stripe magnetic ordering momenta  ($\pi,Q$) and ($Q, \pi$)  ($0<Q\leq\pi/2$)	~\cite{nphys15DFT,prb16Xiang}. 

The real spin state of the nematic phase is mysterious. Based on the LSDA (sGGA) results, one natural speculation is a cooperative paramagnetic state like spin ice~\cite{nature10balents}, i.e. fluctuations within the lowest-energy manifold restore the time-reversal symmetry. It additionally requires that the fluctuations spontaneously condense into either the ($\pi,Q$) or ($Q, \pi$) direction, so the rotation symmetry keeps broken.  Another possibility is that the  LSDA (sGGA) manifold actually collapses into a quantum paramagnetic state like quantum spin liquid~\cite{nphys15Fa}. In this case, the physics is beyond any classical ensemble average.

Given the complexities in the spin degree of freedom, the present work focuses on nonmagnetic calculations only. Practically, we employ GGA instead of sGGA. Given a paramagnetic metal, this is a reasonable starting point. Indeed, for the electronic structure above the nematic transition temperature ($T_s$ $\sim$ 90 K for FeSe), LDA (GGA) instead of LSDA (sGGA) is the common choice to construct the tight-binding model (See, for example, Chpt. 8 in~\cite{book12Dai}). Across $T_s$, the paramagnetic condition of the metallic state does not significantly change~\cite{prb12Mag,ncom16INS}. 

There are two widely-employed recipes to improve the description of strong interactions associated with the Fe $d$-electrons from an orbital-independent mean field to an orbital-dependent potential, which is the key to capture the orbital-ordering instabilities. One recipe is the DFT+$U$ correction~\cite{jpc97LDA+U}, and the other is hybrid functional~\cite{jcp93hybrid}. Successful applications of these corrections to iron-based superconductors were hindered by the observation that +$U$ tends to increase the error of ordered spin moment in the magnetic phase, as already overestimated at the LSDA (sGGA) level. This problem is automatically avoided under the paramagnetic condition. In the absence of the spin degree of freedom, the corrections do not contain intra-orbital (Hubbard) interactions between the spin-up and spin-down electrons, which assist magnetic moment formation. What remains are corrections to the inter-orbital interactions~\cite{jpc97LDA+U}.

A unified view of the +U correction and the hybrid functional under the paramagnetic condition is that they both tend to cancel the unphysical Hartree potential of an electron with itself, which represents one of the most conspicuous error in LDA (GGA), and is significant for a localized orbital. 
While the +$U$ method re-evaluates the interaction terms by projecting $\rho(\textbf{r})$ onto the local orbitals and explicitly exclude the self-interaction term, the hybrid functional incorporates a portion of the exact Fock energy that is also expected to cancel the self-interaction in the Hartree potential.

The main limitation of our calculation is that the short-range AFM correlation and dynamical spin fluctuation in the paramagnetic metal is overlooked, which plays an important role in renormalizing the band width and spectral weight~\cite{nmat11DMFT,nphys11DMFT}.  Nevertheless, it is generally accepted that the mean-field LDA (GGA) treatment nicely reproduces the qualitative features, including the band shapes, orbital components and Fermi surface topology.
Therefore, despite an oversimplified description of paramagnetism, we consider that the symmetric LDA (GGA) band structure is eligible to serve as the numerical parent state, upon which we test whether the residual interaction effects beyond LDA (GGA) drive any instability in the orbital channel.


\section{Results and discussion}
\subsection{New solutions}

\begin{figure}
\centering
\includegraphics[width=8.5cm]{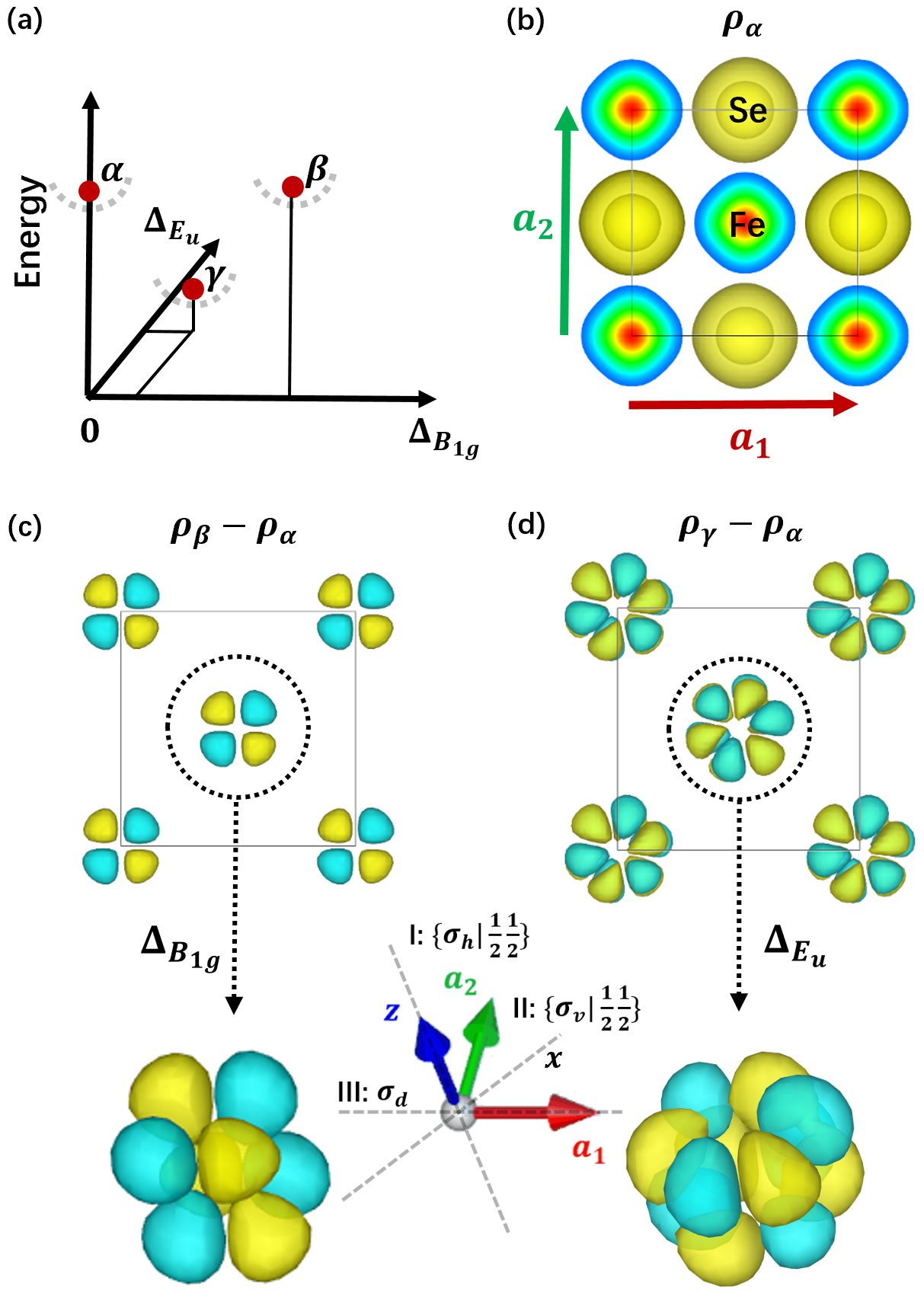}
\caption{(a) Schematic of the three self-consistent solutions in the energy landscape. (b) Charge density contour of the symmetric solution \ ($\rho_\alpha$) with $\mathbf{a}_{1,2}$ as the two lattice vectors of the 2-Fe unit cell; (c) and (d) Charge density change of the two symmetry-breaking solutions ($\delta\rho_\beta$ and $\delta\rho_\gamma$). The blue (yellow) color of the isovalue contour stands for the positive (negative) sign of $\delta \rho$. The zoomed-in distribution of $\delta\rho$ in three dimensions around a single Fe, i.e. the orbital-order parmeters $\Delta_{B_{1g}}$ and  $\Delta_{E_u}$ are displayed below.  The inset shows the three generators of the $P4/nmm / \mathcal{T}\cong {\rm D}_{4h}$ point group. The dashed gray lines are the normals of the reflection mirrors which pass a Fe atom.}
\label{rho}
\end{figure}

With the considerations above, it is reconfirmed that at the plain DFT level only the symmetric solution is obtained. However, including either DFT+$U$ or the hybrid functional, two new solutions stand out, and the symmetric $\rho(\textbf{r})$ becomes a local minimum. Figure \ref{rho}(a) schematically summarizes the three self-consistent solutions: $\alpha$-solution is the symmetric one commonly obtained in previous study; $\beta$ and $\gamma$ are the two symmetry-breaking solutions not known before. Within our calculation framework, we do not identify additional local minima (apart from symmetry-related ones) , and the $\gamma$-solution presents as the global minimum. 

We will first show that the new ground state indeed capture important ARPES observations on the nematic phase, and then come back to analyze the symmetry breaking OPs [Figs. 1(b-d)].

\subsection{Band structure and Fermi surface}\label{sec:compare}

\begin{figure}
\centering
\includegraphics[width=8.5cm]{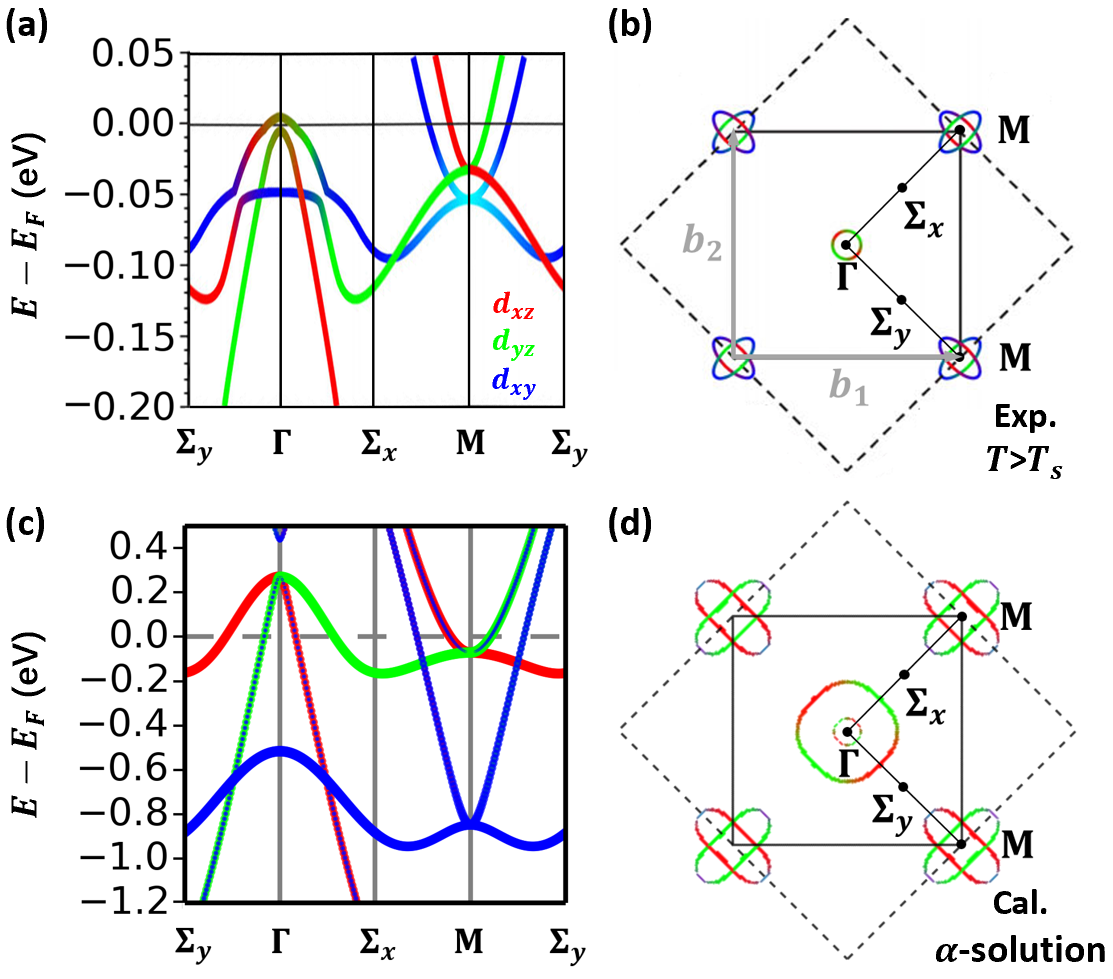}
\caption{The band structure and Fermi surface of the symmetric phase. (a) and (b) are adapted from the ARPES data shown in Ref. \cite{prx19Yi}; (c) and (d) are the calculated bands and $k_z=0$ Fermi surface slices of the symmetric $\alpha$ solution. The color of the bands and Fermi surfaces denotes the orbital components [inset of (a)], following the convention of Ref. \cite{prx19Yi}. $\mathbf{b}_1$ and $\mathbf{b}_2$ are the two reciprocal vectors of the 2-Fe Brillouin zone. The dashed square shows the 1-Fe Brillouin zone, in which $\Gamma$-$\Sigma_x$ ($\Sigma_y$-$\Gamma$) corresponds to the $x$($y$) direction.  }
\label{symmetric}
\end{figure}

\begin{figure}
\centering
\includegraphics[width=8.5cm]{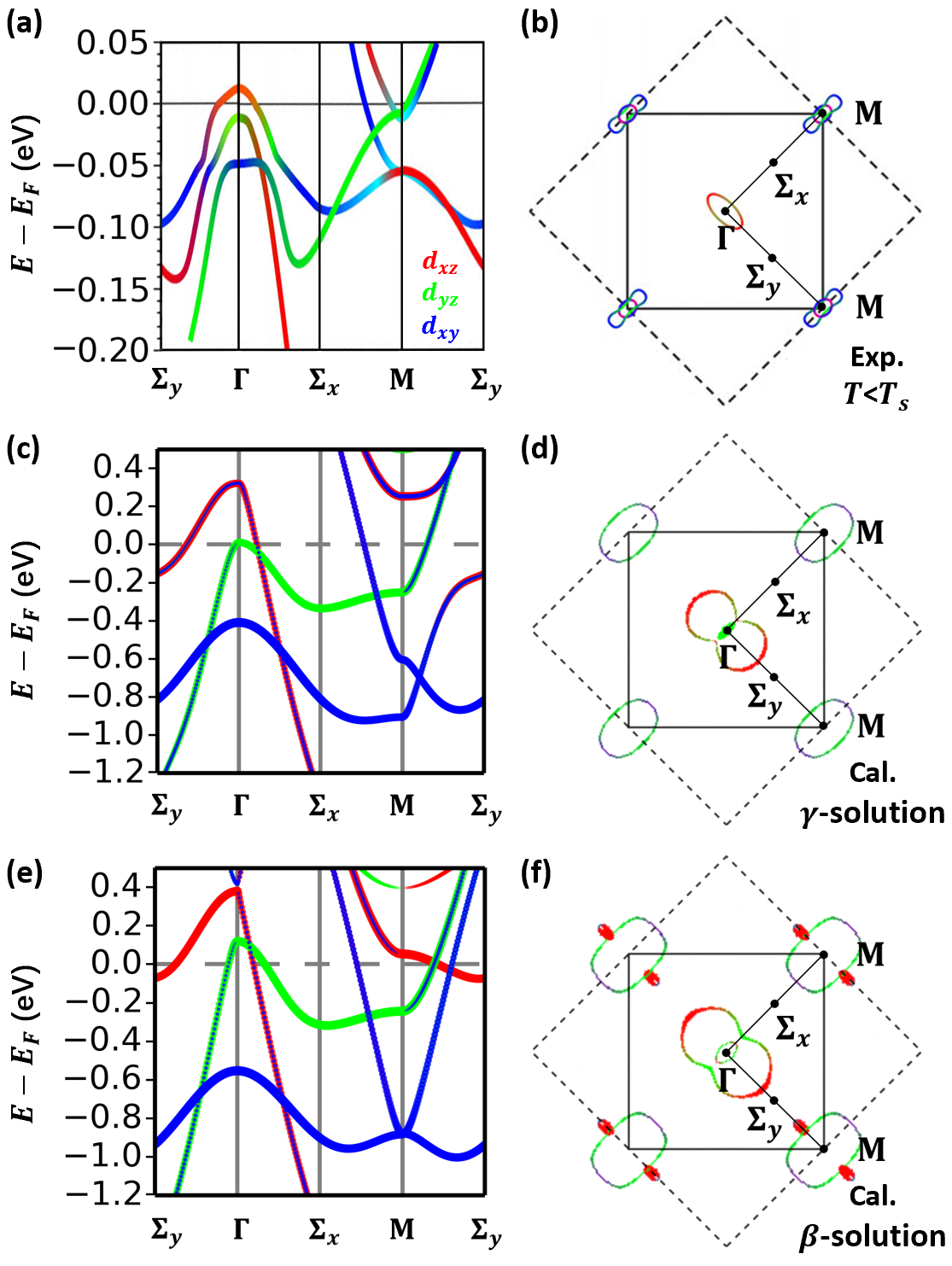}
\caption{The band structure and Fermi surface of the nematic phase. (a) and (b) are adapted from the ARPES data shown in Ref. \cite{prx19Yi}; (c) (d) and (e)(f) are the numerical results of the $\gamma$ and $\beta$ solutions.}
\label{nematic}
\end{figure}

In Fig. \ref{symmetric}, we reproduce the band strucutre and Fermi surface of the symmetric phase, which has been well established in literature~\cite{book12Dai,book15FeSc,rpp11ARPESrev}. The essential features are a hole pocket at $\Gamma$ dominated by $d_{xz}$ and $d_{yz}$, and two elliptical electron pockets intercepted at $M$. The four-lobed shape of the electron pocket arises from a pair of $d_{xz}$ ($d_{yz}$) bands and another pair of $d_{xy}$ bands, which cross the Fermi surface around $M$ [Figs. \ref{symmetric}(a) and (c)]. 

An impressive achievement of DFT is the capability of capturing these features qualitatively, despite the significantly overestimated band width~\cite{qm17review}. Note that the extra inner hole pocket in calculation also relates to the band width error.  Overall, the DFT+$U$ results of the symmetric $\alpha$ solution shown in Figs. \ref{symmetric}(c) and (d) are not much different from the plain DFT ones [See, for example, Chpt. 8 in ~\cite{book12Dai}].  

Turning to the nematic phase, a series of band reconstructions are clearly resolved in the recent ARPES data [Figs. \ref{nematic}(a) and (b)].  Around $\Gamma$, the  $d_{xz}$ and $d_{yz}$ bands are clearly split, and the hole Fermi surface becomes anisotropic. Around $M$, the bands remain largely unchanged on the $\Sigma_x$ side, except for a slight upward shift of the $d_{xz}$ and $d_{yz}$ bands. In contrast, on the $\Sigma_y$ side, the crossing between the $d_{xz}$ and $d_{yz}$ bands is gapped out, regrouping the bands into a hybridized upper branch and a lower branch.  One remarkable consequence is that the  electron Fermi surface loses the lobes along the gapped direction, which is termed as the ``missing electron pocket''~\cite{prx19Yi} , as also observed in \cite{prb15ARPESjp2,prb15ARPESWatson,prb16ARPESWatson,njp17ARPESWatson,prb18ARPESKim,prl16ARPESFeng,prb16ARPESYan,comphy20Korea,sci17QPI,nmat18QPI}. 

It is striking that the $\gamma$ solution as the ground state within our calculation framework captures nearly all the qualitative features. The $\beta$ solution captures the hole band splitting, but it fails to reproduce the characteristic ``one-sided'' gapping. The Fermi surfaces of both the $\gamma$ and $\beta$ solutions clearly indicate that the $C_4$ symmetry is broken [Figs. \ref{nematic}(d)(e)] . It is also the $\gamma$ solution that shows a better agreement with the experimental Fermi surface [Fig. \ref{nematic}(b)]. Naturally, the band width discrepancy is the same as in Fig. \ref{symmetric}.

Another feature of the nematic phase attracting much attention is the momentum dependence of the $d_{xz}$-$d_{yz}$ splitting~\cite{prl16Kontani,prb18Chubukov}. We note that the $\beta$ bands can be largely viewed as a momentum-independent upward shift of the $d_{xz}$ bands.  However, the $\gamma$ bands have a more complicated structure. In particular, around the M point, the originally connected $d_{xz}$ band is gapped into two branches (see the two fractions of red-purple mingled bands around M in Fig. \ref{nematic}(c); the experimental Fig. \ref{nematic}(a) assigns a pure red color to these two branches). If sticking to the lower branch, we can say that the $d_{xz}$-$d_{yz}$ splitting is opposite to the $\Gamma$ point.

\begin{figure}
\centering
\includegraphics[width=8.5cm]{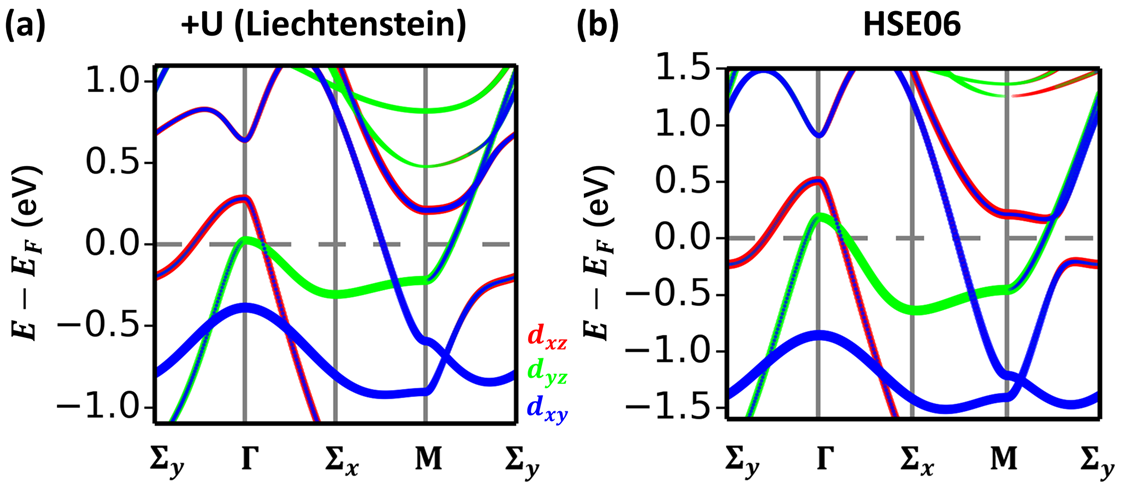}
\caption{The ground-state ($\gamma$ solution) band structure calculated by (a) the full orbital-dependent +$U$ correction as introduced by Liechtenstein et al.~\cite{prb95U}  with $U$=$F_0$=4.8 eV and $J$=($F_2$+$F_4$)/14=1.2 eV and $F_2/F_4$=0.625; and (b) HSE06~\cite{jcp03HSE,jcp06HSE} hybrid functional. }
\label{HSE}
\end{figure}

We note that Figs. 3(c) and (d) are not fine-tuned results. In Fig. \ref{HSE}, we switch to a different version of the +$U$ correction, and benchmark the DFT+$U$ bands to the hybrid functional results. The latter contains no material-specific parameters, and the lattice constants and atomic positions are fully relaxed.  

\subsection{OP analysis}\label{sec:op}

The comparisons between the $\beta$ and $\gamma$ solutions above suggest that $C_4$ symmetry breaking alone is not sufficient to understand the nematic electronic structure observed in ARPES. Distinct from the $\beta$ solution, the $\gamma$ solution should contain some ``hidden orders" unnoticed before. 

Figure 1(b) shows the electron density contour of the $\alpha$ solution [$\rho_\alpha (\textbf{r})$]. In Figs. 1(c) and (d), we subtract $\rho_\alpha (\textbf{r})$ from $\rho_{\beta/\gamma} (\textbf{r})$. The electron density difference [$\delta \rho_{\beta/\gamma} (\textbf{r})$] exhibits clear symmetry breaking patterns. 

$\delta \rho_\beta(\textbf{r})$ can be easily associated with the widely-discussed  ferro-orbital nematic order. It is insightful to revisit Dudarev et al's +$U$ functional~\cite{prb1998Dudarev}. Under the paramagnetic condition, it takes the simple form :
\begin{eqnarray}
E_{+U}=\frac{U}{2} \sum_j ({\rm Tr}\, \tilde{n}_j-{\rm Tr}\, \tilde{n}_j^2),
\end{eqnarray}
in which $j$ labels the Fe site. $\tilde{n}_j$ is the density matrix within the 3$d$-oribtal subspace:
\begin{eqnarray}
\tilde{n}_j^{\alpha\beta}=\sum_i f_i  \langle j\alpha | \psi_i \rangle \langle \psi_i | j\beta \rangle,
\end{eqnarray}
in which $\alpha$ and $\beta$ label the five 3$d$-orbitals of the $j$-th Fe atom. The first term ${\rm Tr}\, \tilde{n}_j$ in the summation is the total occupancy of the 3$d$-orbitals, which is designed to adjust the total energy such that the correction term vanishes if the local orbitals are fully occupied or empty.  This term affects the $p$-$d$ hybridization but does not induce orbital order within the 3$d$-orbitals. The second term $-{\rm Tr}\, \tilde{n}_j^2$ is a self-interaction correction, which can be most easily recognized if $\tilde{n}_j$ is diagonal. In general, this term introduces new OPs in the form of  linear combinations of $\tilde{n}^{\alpha\beta}_j$. 

By inspecting the density matrix within the Fe 3$d$-orbital subspace, we confirm that the most noticeable change is that $\tilde{n}^{yz,yz}-\tilde{n}^{xz,xz}\neq 0$. The $x$,$y$ axes are along the Fe-Fe bonds. Accordingly, the symmetry breaking OP can be written as:
\begin{eqnarray}
 \Delta_{B_{1g}}=\tilde{n}^{yz,yz}-\tilde{n}^{xz,xz} .
\end{eqnarray}
It is known that this OP is Ising type, belonging to the 1D $B_{1g}$ irrep of the D$_{4h}$ point group. Following the choice of the three group generators in Ref.~\cite{prb13group} [See also the inset of Figs. \ref{rho}(c)(d)], we list the transformation matrices (parities for $\Delta_{B_{1g}}$) in Tab. \ref{table-1}.

$\delta \rho_\gamma(\textbf{r})$ in the form of  high-rank multipoles is however  unexpected.   A fraction of $\Delta_{B_{1g}}$ is also present in the $\gamma$ solution. More interestingly, two nonvanishing off-diagonal terms appear: $\tilde{n}^{xz,xy}\simeq\tilde{n}^{yz,x^2-y^2}$ (and the complex conjugate, $\tilde{n}^{xy,xz}\simeq\tilde{n}^{x^2-y^2,yz}$ ).  We define this orbital hybridization as a new OP:
\begin{eqnarray}\label{Eu1}
\Delta_{E_{u,1}}=\tilde{n}^{xz,xy}+\tilde{n}^{yz,x^2-y^2} .
\end{eqnarray}
This is not an Ising OP, which can be most clearly seen by rotating Figs. \ref{rho}(c) and (d) by 90$^{\circ}$. While Fig. \ref{rho}(c) simply changes the sign, Fig. \ref{rho}(d) is transformed into an inequivalent orientation. Formally, the other symmetry-related OP is obtained by interchanging $x$ and $y$:
\begin{eqnarray}\label{Eu2}
\Delta_{E_{u,2}}=\tilde{n}^{yz,xy}-\tilde{n}^{xz,x^2-y^2} .
\end{eqnarray}
These two OPs form a 2D $E_u$ irrep. We list in Tab. \ref{table-1} the associated transformation matrices. 

It is interesting to note that the combination of the two terms in Eqs. (\ref{Eu1}) and (\ref{Eu2}) leads to an analytically compact form of electron-density modulation around Fe atoms, whereas either $\tilde{n}^{xz(yz),xy}$ or $\tilde{n}^{yz(xz),x^2-y^2}$ alone contains a mixture of spherical harmonic functions $Y_l^m$ with ($l=2, m=\pm1$), ($l=4, m=\pm1$) and ($l=4, m=\pm3$). Specifically, 
\begin{eqnarray}
\delta \tilde{\rho}_{E_{u,1}}(\textbf{r})&=&\Delta_{E_{u,1}}[\psi_{xz}(\textbf{r})\psi_{xy}(\textbf{r})+\psi_{yz}(\textbf{r})\psi_{x^2-y^2}(\textbf{r})]\nonumber \\
&\sim&R^2_{3d}(r)\,{\rm sin}^3\theta\,{\rm  cos}\theta\,{\rm sin} 3\phi \nonumber \\
&\sim&Y_{4}^{3}(\theta,\phi)+Y_{4}^{-3}(\theta,\phi),
\end{eqnarray}
in which $R_{3d}(r)$ is the radial wavefunction of 3$d$-orbitals and $(\theta, \phi)$ is the angular coordination of $\textbf{r}$.  Similarly,
\begin{eqnarray}
\delta \tilde{\rho}_{E_{u,2}}(\textbf{r})
&\sim&Y_{4}^{3}(\theta,\phi)-Y_{4}^{-3}(\theta,\phi).
\end{eqnarray}

It is straightforward to check that the analytical $\delta \tilde{\rho}_{E_{u,1}}(\textbf{r})$ indeed reproduces the overall geometry of $\delta \rho_\gamma$ [Fig. \ref{rho}(d)]. It can be viewed as a hexadecapolar order.  

\begin{table}
\caption{Transformation matrices of the predicted OPs with respect to the three generators of the $P4/nmm / \mathcal{T}\cong D_{4h}$ point group. }
\setlength\extrarowheight{3pt}
\begin{ruledtabular}
\begin{tabular}{cccc}
OP & $\{\sigma_h|\frac{1}{2}\frac{1}{2}\}$  & $\{\sigma_v|\frac{1}{2}\frac{1}{2}\}$ & $\sigma_d$\\
\hline
$\Delta_{B_{1g}}$ & 1 & 1& -1 \\
$(\Delta_{E_{u,1}},\Delta_{E_{u,2}})$ & $\left(\protect\begin{array}{cc}
  1 & 0    \protect\\
  0 & 1 
\protect\end{array}\right)$
 &  $\left(\protect\begin{array}{cc}
  -1 & 0    \protect\\
  0 & 1 
\protect\end{array}\right)$ &  $\left(\protect\begin{array}{cc}
  0 & 1    \protect\\
  1 & 0 
\protect\end{array}\right)$ 

\end{tabular}
\end{ruledtabular}
\label{table-1}
\end{table}

\subsection{Experimental consequences}

\begin{table}
\caption{Point groups of the three solutions at $\Gamma$ and along $\Sigma_{x(y)}$ directions.}
\setlength\extrarowheight{3pt}
\begin{ruledtabular}
\begin{tabular}{cccc}
Solution & $\Gamma$  & $\Sigma_x$ & $\Sigma_y$\\
\hline
$\alpha$ & D$_{4h}$ & C$_{2v}$ & C$_{2v}$ \\
$\beta$ & D$_{2h}$ & C$_{2v}$ & C$_{2v}$ \\
$\gamma$ & C$_{2v}$ & C$_{2v}$ & C$_{s}$ \\

\end{tabular}
\end{ruledtabular}
\label{table-2}
\end{table}

\subsubsection{Band reconstructions}
Identification of the OPs reveal the physical origin of band reconstructions shown in Fig. \ref{nematic}. According to Tab. \ref{table-1}, both $\Delta_{B_{1g}}$ and $\Delta_{E_u}$ breaks $\sigma_d$, and thus the $C_4$ rotation, leading to nematicity. The hole band splitting is primarily associated with this symmetry breaking, and the $\beta$ and $\gamma$ solutions behave similarly. 

The``hidden order" in $\Delta_{E_u}$ is related to the further breaking of $\{{\sigma_{v}|\frac{1}{2}\frac{1}{2}}\}$.  The remaining symmetry shrinks to a C$_{2v}$ group. We list in Tab. \ref{table-2} the little groups of the three solutions. It is important to notice that $\Delta_{B_{1g}}$ does not lower the symmetry of the ordinary k-points along $\Sigma_{x(y)}$ directions. However, $\Delta_{E_u}$ lowers the symmetry along $\Sigma_y$ from C$_{2v}$ to C$_s$. This missing group element originally protects the $d_{xz}$ and $d_{xy}$ band crossing near the $M$ point. This symmetry reason naturally explains why the ``one-sided" gapping occurs, without the requirement of band inversion~\cite{prx19Yi} or orbital-selective quasiparticle weight~\cite{sci17QPI,nmat18QPI} as previously proposed.

From a different angle, the ``one-sided" gapping can be understood from the first term on the right hand side of Eq. (\ref{Eu1}). This hybridization term tends to gap crossing points between $d_{xz}$ and $d_{xy}$ bands. On the other hand, $d_{yz}$ is coupled to $d_{x^2-y^2}$ only, which lies deep below the Fermi surface. The crossing point between $d_{yz}$ and $d_{xy}$ bands is thus largely unperturbed. The redistribution of electron population between the $d_{xz}$, $d_{yz}$ and $d_{xy}$ orbitals below $T_s$ was recently noticed in experiment~\cite{prx20NMR}. 

\subsubsection{Inversion symmetry breaking}\label{sec:exp}
Another important difference between the C$_{2v}$ and the  D$_{2h}$ point group is that C$_{2v}$ does not contain inversion symmetry. Direct measurement sensitive to electronic inversion symmetry breaking is now possible, thanks to the second harmonic generation (SHG) technique~\cite{nphys17SHG,prb16SHG,acta12SHG}. The prediction is that if the $\Delta_{E_{u,1(2)}}$ OP occurs, an unambiguous SHG signal should be observed below $T_s$. 

$\Delta_{E_{u,1(2)}}$ is conjugate to an in-plane electric field $\vec{E}$ along one of the the Fe-Fe bonding directions. This will lead to a Dresselhaus splitting $\sim(\vec{E}\times\vec{k})\cdot \vec{S}$, in which $\vec{S}$ is the electron spin operator. The splitting is largest along the k-direction perpendicular to $\vec{E}$, and vanishes when $\vec{k}\parallel\vec{E}$. This term also leads to an out-of-plane polarization of the Fe spins, which is consistent with the spin-polarized inelastic neutron scattering data~\cite{prx17SOC}.

\begin{figure}
\centering
\includegraphics[width=15cm]{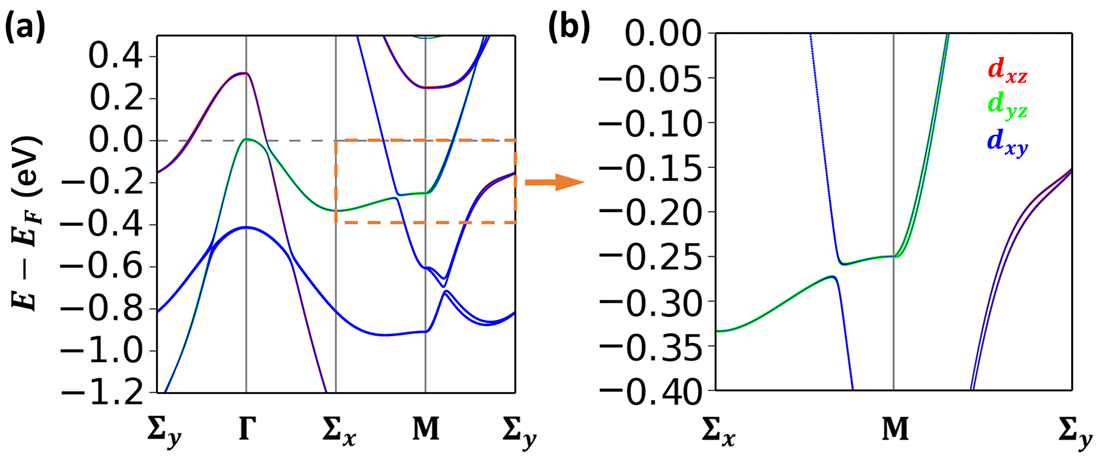}
\caption{(a) Ground-state ($\gamma$ solution) band structure including SOC; (b) One-sided Dresselhaus splitting around the electron pocket. }
\label{soc}
\end{figure}

We quantify the Dresselhaus splitting by acting spin-orbit coupling (SOC) upon the paramagnetic $\gamma$ solution as a first-order perturbation. Due to the overestimated band width, the SOC effect can be barely observed in Fig. \ref{soc}(a), but a zoomed-in view [Fig. \ref{soc}(b)] clearly shows the directional splitting of the order of $O(10)$ meV. This is also considered as a unique signature of $\Delta_{E_u}$, which can in principle be resolved in high-precision ARPES data, e.g. Ref.~\cite{arxiv19ARPES},  and the Supplementary Material of Ref.~\cite{prb18ARPESBorisenko}.

In the same way as $\Delta_{B_{1g}}$ couples to a shear strain of the lattice~\cite{nphys14nem}, $\Delta_{E_u}$ also couples to a special type of structural distortion. According to the full relaxation at the hybrid functional level,  a relative glide between the Fe and Se layers occur along one of the Fe-Fe bonding directions, which also breaks the inversion symmetry in analogy to the polar distortion in ferroelectric materials. However, the calculated magnitude is as small as 0.001$\rm{\AA}$, which again reflects that lattice instability is not the driving force. 

\subsection{Dependence on microscopic details}

\begin{figure*}
\centering
\includegraphics[width=15cm]{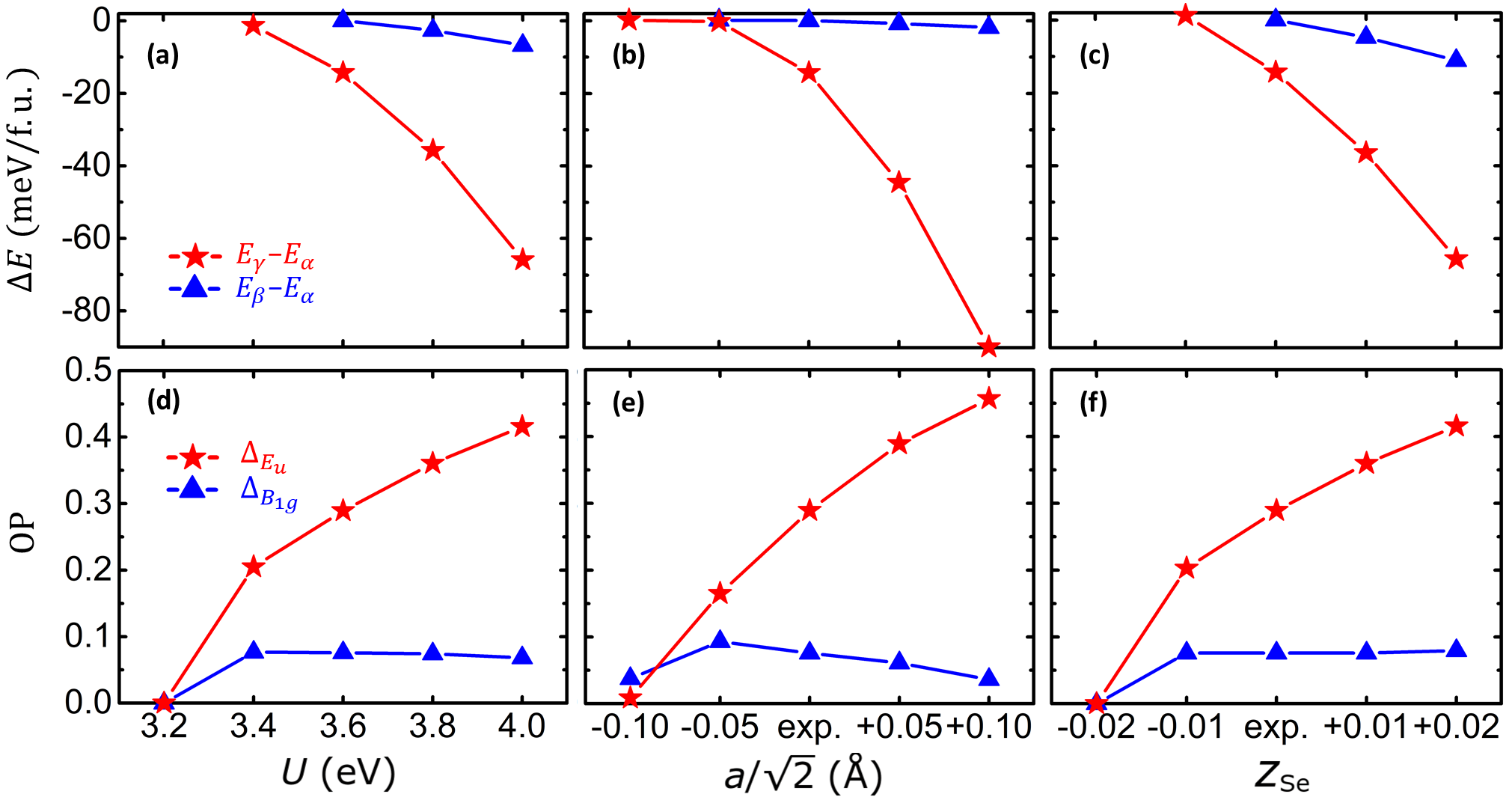}
\caption{ The evolution of (a-c)the total energy of the $\beta$ (blue triangles) and $\gamma$ (red stars) solutions; and (d-f) OP amplitude $\Delta_{E_u}$ (red stars) and $\Delta_{B_{1g}}$ (blue triangles) of the ground state ($\gamma$ solution) as a function of the the +$U$ parameter ($U$), Fe-Fe distance ($a/\sqrt{2}$) and the out-of-plane coordinate of Se ($Z_{Se}$).}
\label{opevolve}
\end{figure*}

\begin{figure*}
\centering
\includegraphics[width=17cm]{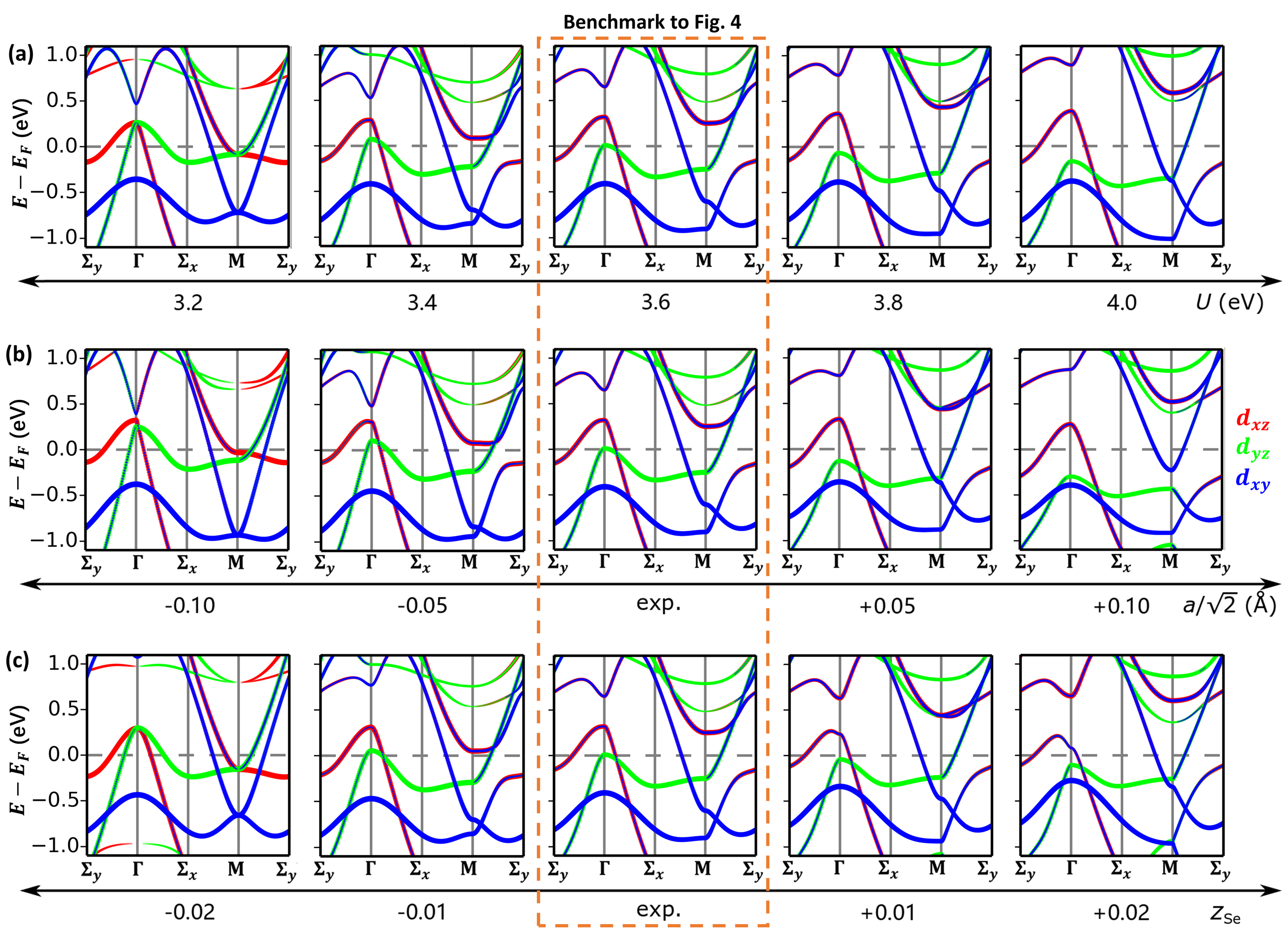}
\caption{ The evolution of the ground-state band structure as a function of the +$U$ parameter $U$), Fe-Fe distance ($a/\sqrt{2}$) and the out-of-plane coordinate of Se ($Z_{Se}$). }
\label{bandevolve}
\end{figure*}

Figure \ref{opevolve} shows how the total energy of the three solutions change when we manually change the +$U$ parameter ($U$), Fe-Fe distance ($a/\sqrt{2}$) and the out-of-plane coordinate of Se ($Z_{Se}$). The overall trend is that nematicity is favored by (a) a larger U (interaction driven) and (b) larger $a/\sqrt{2}$ and $Z_{Se}$, which coincides with the fact that nematicity is suppressed by pressure~\cite{ncom16highp}. Compared with $a/\sqrt{2}$, $Z_{Se}$ appears as a more sensitive factor. $\Delta_{E_u}$ and $\Delta_{B_{1g}}$ typically coexists in the ground state, but $\Delta_{E_u}$ is the leading one. The existence of the $E_u$ order parameters will generically generate the $B_{1g}$ order parameter, due to the symmetry-allowed coupling $\Delta_{B1g} (\Delta_{Eu,1}^2-\Delta_{Eu,2}^2)$ in an effective Ginzburg-Landau-type theory. The pure  $\Delta_{B_{1g}}$ ($\beta$ solution) always has a higher energy. 

Figure \ref{bandevolve} summarizes the ground-state band evolution as we change the microscopic details. It is informative to observe how the nematic band structure deforms back to the symmetric one on the left side of the figure. During the process, the $d_{xz}$-$d_{xy}$ gap along the $M$-$\Sigma_y$ direction gradually closes. The linear crossing without hybridization forms in the end.

\section{Conclusion}
We demonstrate a first-principles approach to reproduce the paramagnetic nematic state in FeSe, without breaking either the tetragonal lattice symmetry or the time-reversal symmetry. We incorporate orbital-resolved interactions by +$U$ and hybrid functional, and precondition the initial wavefunction to find self-consistent solutions with spontaneous symmetry breaking. The lowest-energy nematic state we find features a two-component vector OP belonging to the $E_u$ irrep, in addition to the Ising ferro-orbital order, which is important to produce the correct Fermi surface topology. We propose that the inversion symmetry breaking induced by the $E_u$ OP can be detected by high-precision measurement of the band dispersion as well as SHG.

\section{Methods}

\subsection{Rountine setup}
All the calculations are performed with respect to bulk FeSe by using the Vienna $ab$ $initio$ Simulation Package~\cite{prb96VASP}. The lattice structure keeps tetragonal throughout the DFT and DFT+$U$ study with the full $P4/nmm$ space group symmetry. The automatic symmetrisation routine is switched off to probe electronic spontaneous symmetry breaking. The experimental lattice parameters~\cite{prb09structure} are used as the reference to obtain the main results. The atomic positions are relaxed until the residual forces are smaller than  $5\times10^{-3}$ eV/$\rm\AA$. As a benchmark, we also perform full structural relaxation using the hybrid functional. In the end , the in-plane lattice constant and the Se height are manually varied separately to understand their effects.  

The plane-wave cutoff is 500 eV  in combination with the projector augmented wave method~\cite{prb94PAW}. The Monkhorst-Pack~\cite{prb76MP} k-point grid is $12\times12\times6$.  The exchange and correlation is treated by using the Perdew-Burke-Ernzerh generalized gradient approximation (GGA) functional~\cite{prl96PBE}. The convergence criteria is $10^{-6}$ eV for electronic iterations. 

Unless specified otherwise, the presented calculation results include the rotational invariant +$U$ correction introduced by Dudarev et al.~\cite{prb1998Dudarev} with $U$=3.6 eV (or more rigorously, $U$-$J$=3.6 eV). This +$U$ parameter is selected by benchmarking the band structure to HSE06~\cite{jcp03HSE,jcp06HSE} hybrid functional results, which contains no material-specific parameter. The more complicated orbital-dependent +$U$ correction as introduced by Liechtenstein et al.~\cite{prb95U} with $U$=$F_0$=4.8 eV and $J$=($F_2$+$F_4$)/14=1.2 eV and $F_2/F_4$=0.625 gives very similar results. In the end, we purposely tune Dudarev's +$U$ parameter from 3.2 to 4.0 eV to understand its effect on the results.   

\subsection{Wavefunction preconditioning}\label{sec:precon}

One important numerical issue is that for a correlated system like FeSe, the energy landscape could be rather complicated. In consequence, numerical minimization might be easily trapped to some local minimum points~\cite{prb05FeO}. In particular, starting from an initial electron density $\rho_0 (\textbf{r})$ that respects the full symmetry of the underlying lattice, the iteration can easily end at a $\rho_{GS} (\textbf{r})$ without symmetry breaking, if a symmetric local minimum exists. 
Therefore, to probe potential symmetry breaking,  it is beneficial to purposely drag $\rho_0 (\textbf{r})$ away from the symmetric basin by preconditioning the initial trial wavefunctions $\{\psi_i (\textbf{r})\}$. 

To test whether the nematic order can spontaneously develop, we first generate a set of ${\psi_i (\textbf{r})}$ from a preparatory calculation on a manually distorted FeSe lattice that slightly breaks the symmetry. Then this set of $\{\psi_i (\textbf{r})\}$ is fed to an undistorted FeSe lattice as a starting point to see whether it flows to a different local minimum. Specifically,  we find that preconditioned wavefunctions generated by an uniaxial strain along either the $x$ or $y$ axis (see the inset of Fig. \ref{rho}) tend to flow into the $\beta$ ($B_{1g}$) basin. For the $\gamma$ ($E_u$) solution, the most convenient preconditioning is to shift the origin of the Fe layer by a small amount along either the $x$ or the $y$ axis. We note that preconditioning is merely a numerical treatment to better search the complicated energy landscape. It does not change the landscape. In other words, no matter how the initial $\{\psi_i (\textbf{r})\}$ is preconditioned, the electronic Hamiltonian for the production run is always ensured to be an invariant of the space group $P4/nmm$. 

It is worth mentioning that besides wavefunction preconditioning, the minimization algorithm is also a matter of concern. We notice that the damped velocity friction algorithm (for electronic minimization) sometimes has a better performance of escaping a shallow local minimum than the blocked Davidson iteration scheme and the direct inversion in the iterative subspace scheme~\cite{prb96VASP}. Without wavefunction preconditioning, the damped velocity friction algorithm can still correctly find the $\gamma$ ($E_u$) solution as the ground state, despite a significantly larger number of iterations.

\section{Data availability}
The data that support the findings of this study are available from the corresponding author upon reasonable request.

\section{Acknowledgement}
We would like to thank Ming Yi, Yan Zhang, Wei Li, Yuan Li, Tao Wu, Yi Zhou, Yuan Wan, Hong Yao and Yuanming Lu for helpful discussion. This work is supported by NSFC under Grant No. 11774196 and Tsinghua University Initiative Scientific Research Program. FW acknowledges support from the National Key Research and Development Program of China (Grand No. 2017YFA0302904).

\section{Author contributions}
X. L and Z. L. conceived the project. X. L. and S. Z. performed the calculations. F. W contributed to the theoretical interpretations. All the authors prepared the manuscript. 

\section{Additional information}
Correspondence should be addressed to Z.L.

\section{Competing financial interests}
The authors declare no competing financial interests.

\end{document}